\documentclass[a4paper]{eptcs}
\usepackage[english]{babel}
\usepackage{hyperref}
\usepackage{appendix}
\usepackage{listings}
\usepackage[usenames,dvipsnames]{color}
\usepackage{graphicx}
	\DeclareGraphicsExtensions{.pdf,.png,.jpg,.jpeg,.mps,.emf}
\usepackage{epstopdf}
\usepackage{multirow}
\usepackage{xspace}	
\usepackage{mdwlist}
\usepackage{prooftree}
\usepackage{amsmath, amsthm, amssymb,xcolor,wasysym, altbeginend}
\usepackage{comment}		
\usepackage{mathptmx,latexsym }
\usepackage{stmaryrd}
\usepackage{eurosym}
\usepackage{tabularx}
\usepackage{wrapfig}
\usepackage[font=small,labelfont=bf, justification=centering]{caption}
\usepackage{breakurl}
\usepackage{subcaption}
\usepackage{enumitem}
\setlist{noitemsep,topsep=0pt,parsep=0pt,partopsep=0pt}
\setenumerate{noitemsep,topsep=0pt,parsep=0pt,partopsep=0pt}

\hypersetup{
  pdftitle={Refactoring Tool for F\#},
  pdfauthor={Authors all over the place},
  pdfsubject={Refactoring Tool for F\#},
  pdfkeywords={fsharp, documentation, searching},
  colorlinks=true,
  linkcolor	= black, 
  citecolor = black,
  urlcolor = black
}

\lstdefinelanguage{FSharp}{
  morekeywords={let, fun, for, mutable, if, then, else, match, with, of, type, abstract, member, override, and,  interface},
  otherkeywords={=>,<-,<\%,<:,>:,\#,@, true, false, ->},
  sensitive=true,
  morecomment=[l]{//},
  morestring=[b]",
  morestring=[b]',
  morestring=[b]"""
}

\lstdefinelanguage{xml}
{
  morestring=[b]",
  morecomment=[s]{<?}{?>},
  stringstyle=\color{RoyalBlue},
  identifierstyle=\color{BrickRed},
  keywordstyle=\color{red},
  morekeywords={xmlns,value,guid, type, name, id}
}

\begin{document}
%

%

%
%
%
%
%
%

\makeatletter

%
\let\if@envcntreset\iffalse
\let\if@envcntsame\iffalse
\let\if@envcntsect\iftrue

%
\def\@thmcountersep{}
\def\@thmcounterend{.}

\def\spnewtheorem{\@ifstar{\@sthm}{\@Sthm}}


\def\@spnthm#1#2{%
  \@ifnextchar[{\@spxnthm{#1}{#2}}{\@spynthm{#1}{#2}}}
\def\@Sthm#1{\@ifnextchar[{\@spothm{#1}}{\@spnthm{#1}}}

\def\@spxnthm#1#2[#3]#4#5{\expandafter\@ifdefinable\csname #1\endcsname
   {\@definecounter{#1}\@addtoreset{#1}{#3}%
   \expandafter\xdef\csname the#1\endcsname{\expandafter\noexpand
     \csname the#3\endcsname \noexpand\@thmcountersep \@thmcounter{#1}}%
   \expandafter\xdef\csname #1name\endcsname{#2}%
   \global\@namedef{#1}{\@spthm{#1}{\csname #1name\endcsname}{#4}{#5}}%
                              \global\@namedef{end#1}{\@endtheorem}}}

\def\@spynthm#1#2#3#4{\expandafter\@ifdefinable\csname #1\endcsname
   {\@definecounter{#1}%
   \expandafter\xdef\csname the#1\endcsname{\@thmcounter{#1}}%
   \expandafter\xdef\csname #1name\endcsname{#2}%
   \global\@namedef{#1}{\@spthm{#1}{\csname #1name\endcsname}{#3}{#4}}%
                               \global\@namedef{end#1}{\@endtheorem}}}

\def\@spothm#1[#2]#3#4#5{%
  \@ifundefined{c@#2}{\@latexerr{No theorem environment `#2' defined}\@eha}%
  {\expandafter\@ifdefinable\csname #1\endcsname
  {\global\@namedef{the#1}{\@nameuse{the#2}}%
  \expandafter\xdef\csname #1name\endcsname{#3}%
  \global\@namedef{#1}{\@spthm{#2}{\csname #1name\endcsname}{#4}{#5}}%
  \global\@namedef{end#1}{\@endtheorem}}}}

\def\@spthm#1#2#3#4{\topsep 7\p@ \@plus2\p@ \@minus4\p@
\refstepcounter{#1}%
\@ifnextchar[{\@spythm{#1}{#2}{#3}{#4}}{\@spxthm{#1}{#2}{#3}{#4}}}

\def\@spxthm#1#2#3#4{\@spbegintheorem{#2}{\csname the#1\endcsname}{#3}{#4}%
                    \ignorespaces}

\def\@spythm#1#2#3#4[#5]{\@spopargbegintheorem{#2}{\csname
       the#1\endcsname}{#5}{#3}{#4}\ignorespaces}

\def\@spbegintheorem#1#2#3#4{\trivlist
                 \item[\hskip\labelsep{#3#1\ #2\@thmcounterend}]#4}

\def\@spopargbegintheorem#1#2#3#4#5{\trivlist
      \item[\hskip\labelsep{#4#1\ #2}]{#4(#3)\@thmcounterend\ }#5}


\def\@sthm#1#2{\@Ynthm{#1}{#2}}

\def\@Ynthm#1#2#3#4{\expandafter\@ifdefinable\csname #1\endcsname
   {\global\@namedef{#1}{\@Thm{\csname #1name\endcsname}{#3}{#4}}%
    \expandafter\xdef\csname #1name\endcsname{#2}%
    \global\@namedef{end#1}{\@endtheorem}}}

\def\@Thm#1#2#3{\topsep 7\p@ \@plus2\p@ \@minus4\p@
\@ifnextchar[{\@Ythm{#1}{#2}{#3}}{\@Xthm{#1}{#2}{#3}}}

\def\@Xthm#1#2#3{\@Begintheorem{#1}{#2}{#3}\ignorespaces}

\def\@Ythm#1#2#3[#4]{\@Opargbegintheorem{#1}
       {#4}{#2}{#3}\ignorespaces}

\def\@Begintheorem#1#2#3{#3\trivlist
                           \item[\hskip\labelsep{#2#1\@thmcounterend}]}

\def\@Opargbegintheorem#1#2#3#4{#4\trivlist
      \item[\hskip\labelsep{#3#1}]{#3(#2)\@thmcounterend\ }}

\if@envcntsect
   \def\@thmcountersep{.}
   \spnewtheorem{theorem}{Theorem}[section]{\bfseries}{\itshape}
\else
   \spnewtheorem{theorem}{Theorem}{\bfseries}{\itshape}
   \if@envcntreset
      \@addtoreset{theorem}{section}
   \else
      \@addtoreset{theorem}{chapter}
   \fi
\fi

\spnewtheorem*{claim}{Claim}{\itshape}{\rmfamily}
\if@envcntsame 
   \def\spn@wtheorem#1#2#3#4{\@spothm{#1}[theorem]{#2}{#3}{#4}}
\else 
   \if@envcntsect 
      \def\spn@wtheorem#1#2#3#4{\@spxnthm{#1}{#2}[section]{#3}{#4}}
   \else 
      \if@envcntreset
         \def\spn@wtheorem#1#2#3#4{\@spynthm{#1}{#2}{#3}{#4}
                                   \@addtoreset{#1}{section}}
      \else
         \def\spn@wtheorem#1#2#3#4{\@spynthm{#1}{#2}{#3}{#4}
                                   \@addtoreset{#1}{chapter}}%
      \fi
   \fi
\fi
\spn@wtheorem{case}{Case}{\itshape}{\rmfamily}
\spn@wtheorem{conjecture}{Conjecture}{\itshape}{\rmfamily}
\spn@wtheorem{corollary}{Corollary}{\bfseries}{\itshape}
\spn@wtheorem{definition}{Definition}{\bfseries}{\itshape}
\spn@wtheorem{example}{Example}{\itshape}{\rmfamily}
\spn@wtheorem{exercise}{Exercise}{\itshape}{\rmfamily}
\spn@wtheorem{lemma}{Lemma}{\bfseries}{\itshape}
\spn@wtheorem{problem}{Problem}{\itshape}{\rmfamily}
\spn@wtheorem{property}{Property}{\itshape}{\rmfamily}
\spn@wtheorem{proposition}{Proposition}{\bfseries}{\itshape}
\spn@wtheorem{question}{Question}{\itshape}{\rmfamily}
\spn@wtheorem{solution}{Solution}{\itshape}{\rmfamily}
\spn@wtheorem{remark}{Remark}{\itshape}{\rmfamily}

\def\@takefromreset#1#2{%
    \def\@tempa{#1}%
    \let\@tempd\@elt
    \def\@elt##1{%
        \def\@tempb{##1}%
        \ifx\@tempa\@tempb\else
            \@addtoreset{##1}{#2}%
        \fi}%
    \expandafter\expandafter\let\expandafter\@tempc\csname cl@#2\endcsname
    \expandafter\def\csname cl@#2\endcsname{}%
    \@tempc
    \let\@elt\@tempd}

\def\theopargself{\def\@spopargbegintheorem##1##2##3##4##5{\trivlist
      \item[\hskip\labelsep{##4##1\ ##2}]{##4##3\@thmcounterend\ }##5}
                  \def\@Opargbegintheorem##1##2##3##4{##4\trivlist
      \item[\hskip\labelsep{##3##1}]{##3##2\@thmcounterend\ }}
      }

\makeatother

\newcommand{\MYPARAGRAPH}[1]{\paragraph{\textnormal{\textbf{#1}}}}

\definecolor{dkblue}{rgb}{0,0.1,0.5}
\definecolor{dkgreen}{rgb}{0,0.4,0}
\definecolor{dkred}{rgb}{0.4,0,0}

\newcommand{\CODESIZE}{\small}
\newcommand{\CODESTYLE}{\ttfamily}

\lstset%
{%
	captionpos=b,
	columns=flexible, 
	commentstyle=\CODESTYLE\footnotesize\color{purple},
	escapeinside={*@}{@*},
	float=hbp,
	frame=none,
	language=Java,
	mathescape=true,
	numbers=none, 
	numberstyle=\tiny,
	showspaces=false,
	showstringspaces=false,
	showtabs=false,
	stringstyle=\color{teal},
	tabsize=2
}

\newcommand{\CODE}[1]{\texttt{\CODESIZE#1}} 
\newcommand{\GREYCODE}[1]{\CODE{
	\color{black}{#1}}} 
\newcommand{\SJCODE}[1]
{%
	\lstinline[style=SJ]+#1+%
}

\lstnewenvironment{CODELISTING}
{%
	\onehalfspacing
	\lstset%
	{%
		basicstyle=\CODESTYLE\footnotesize,
		keywordstyle=\CODESTYLE\footnotesize,
	}
}%
{%
	\doublespacing 
}

\lstdefinestyle{SJ}%
{%
	basicstyle   = \CODESTYLE\footnotesize,
	keywordstyle=[1]{
	\!\!\!\color{dkblue}
	\CODESTYLE\footnotesize},
	keywordstyle=[2]{
	\!\!\!\color{dkgreen}
	\CODESTYLE\footnotesize},
        moredelim=*[s][\footnotesize\color{dkgreen}]{<}{>},
        morekeywords =
	[1]{%
		protocol, role, choice, at, or, from, to, rec, parallel,
                and, interrupt, by, finish, continue, global, local, self
	},
	morekeywords =
	[2]{%
                int,Data,
	},
       literate={>=}{$\geq\ $}{2}{<=}{$\leq\ $}{2}
}

\lstnewenvironment{SJLISTING}%
{
	\lstset{style=SJ}
}
{
}

%
\newcommand{\REF}[1]{\S\,\ref{#1}}

%
\newcommand{\OPASSIGN}{\, \CODE{:=} \,}
\newcommand{\OPEQ}{\ensuremath{=}}
\newcommand{\OPDEC}{\CODE{--}}
\newcommand{\OPINC}{\CODE{++}}
\newcommand{\OPSEQ}{\CODE{;}}
\newcommand{\OPNOT}{\ensuremath{\neg}}

\newcommand{\OLINE}[1]{\ensuremath{\overline{#1}}}  
\newcommand{\SET}[1]{\ensuremath{\{ #1 \}}}         
\newcommand{\FUN}[2]{\ensuremath{\mathsf{#1}(#2)}}  

%
\newcommand{\KWORD}[1]{\ensuremath{\mathsf{#1}}}
\newcommand{\DTYPE}[1]{\ensuremath{\mathtt{#1}}}  
\newcommand{\DVAL}[1]{\ensuremath{\mathtt{#1}}}   
\newcommand{\PPAR}{\ensuremath{\, | \,}}
\newcommand{\PPARGROUP}[1]{\ensuremath{(#1)}}
\newcommand{\PIF}{\ensuremath{\KWORD{if}}}
\newcommand{\PTHEN}{\ensuremath{\KWORD{then}}}
\newcommand{\PELSE}{\ensuremath{\KWORD{else}}}
\newcommand{\PNIL}{\ensuremath{\mathbf{0}}}
\newcommand{\EMPTY}{\ensuremath{\epsilon}}

%
\newcommand{\LAB}[1]
	{\ensuremath{#1}}            
\newcommand{\LABVAL}[1]
	{\ensuremath{\mathtt{#1}}}   
\newcommand{\ROLE}[1]
	{\participant{#1}}           
\newcommand{\PARTY}[1]
	{\ensuremath{\mathsf{#1}}}   
\newcommand{\MSGlxS}[3]
	{\ensuremath{\MSGlx{#1}{#2\!:\!#3}}}  
\newcommand{\MSGlx}[2]
	{\ensuremath{#1 (#2)}}              
\newcommand{\MSGlS}[2]
	{\MSGlx{#1}{#2}}                    
\newcommand{\BRANCH}[1]
	{\ensuremath{\SET{#1}}}         
\newcommand{\MUREC}[1]
	{\ensuremath{\mu \RECVAR{#1}}}  
\newcommand{\RECVAR}[1]
	{\ensuremath{\keyword{#1}}}     

%
\newcommand{\STATEVAR}[1]
	{\ensuremath{\mathtt{#1}}}                
\newcommand{\ROLEVAR}[2]
	{\ensuremath{\ROLE{#1} . \STATEVAR{#2}}}  
\newcommand{\STATEVARDECL}[2]
	{\ensuremath{\STATEVAR{#1} : \DTYPE{#2}}}      
\newcommand{\PARTYSTATEDECL}[2]
	{\ensuremath{\ROLE{#1}} : [#2]}                

%
\newcommand{\GSEP}{\ensuremath{.}}  
\newcommand{\GLOBAL}[1]
	{\ensuremath{\mathcal{#1}}}       
\newcommand{\GLOBALi}[2]
	{\ensuremath{\GLOBAL{#1}_{#2}}}   
\newcommand{\GSEND}[2]
	{\ensuremath{\ROLE{#1} \rightarrow \ROLE{#2} :}}  

\newcommand{\GBRA}[1]
	{\ensuremath{\SET{#1}}}         
\newcommand{\GREC}[1]
	{\ensuremath{\mu \RECVAR{#1}}}  

%
\newcommand{\GLOBALDECL}[1]
	{\ensuremath{((#1))}}             
\newcommand{\LASS}[1]
	{\ensuremath{\langle #1 \rangle}}  
\newcommand{\LEFF}[1]
	{\LASS{#1}}                        
\newcommand{\LASSEFF}[2]
	{\ensuremath{\LASS{#1,\, #2}}}     
\newcommand{\RASS}[1]
	{\ensuremath{\{ #1 \}}}            
\newcommand{\REFF}[1]
	{\RASS{#1}}                        
\newcommand{\RASSEFF}[2]
	{\ensuremath{\RASS{#1,\, #2}}}     
\newcommand{\GRECtexA}[4]
	{\ensuremath{\MUREC{#1} \langle #2 \rangle (#3) \{ #4 \}}}
\newcommand{\GRECVARte}[2]
	{\ensuremath{\RECVAR{#1} \langle #2 \rangle}}

%
\newcommand{\LSEP}{\ensuremath{.}}  
\newcommand{\LOCAL}[1]
	{\ensuremath{\mathcal{#1}}}       
\newcommand{\LOCALi}[2]
	{\ensuremath{\LOCAL{#1}_{#2}}}    
\newcommand{\LSEND}[1]
	{\ensuremath{\ROLE{#1} \,!\,}}    
\newcommand{\LRECV}[1]
	{\ensuremath{\ROLE{#1} \,?}}      

%
\newcommand{\LOCALDECL}[1]
	{\ensuremath{[#1]}}               
\newcommand{\LRECtexA}[4]
	{\GRECtexA{#1}{#2}{#3}{#4}}
\newcommand{\LRECVARte}[2]
	{\GRECVARte{#1}{#2}}

%
\newcommand{\POSEP}{\ensuremath{;}}  
\newcommand{\PESEP}{\ensuremath{;}}  
\newcommand{\PSEP}{\ensuremath{.}}   
\newcommand{\PINIT}[1]
	{\ensuremath{\mathsf{#1}}}         
\newcommand{\PINITi}[2]
	{\ensuremath{\mathsf{#1}_{#2}}}    
\newcommand{\PREQ}[4]
	{\ensuremath{\OLINE{#1} \langle #2 [\ROLE{#3}] : \GLOBAL{#4} \rangle}}
\newcommand{\PACC}[4]
	{\ensuremath{#1 ( #2 [ \ROLE{#3} ] : \GLOBAL{#4} )}}          
\newcommand{\PSEND}[5]
	{\ensuremath{#1 [\ROLE{#2}, \ROLE{#3}] \,!\, \LAB{#4} \langle #5 \rangle}}
\newcommand{\PRECV}[3]
	{\ensuremath{#1 [\ROLE{#2}, \ROLE{#3}] \,?\,}}                
\newcommand{\PRECX}[1]
	{\ensuremath{\mu #1}}                            
\newcommand{\PRECXx}[2]
	{\ensuremath{\PRECX{#1} (#2)}}                   
\newcommand{\PRECe}[1]
	{\ensuremath{\langle #1 \rangle}}                
\newcommand{\PRECVARX}[1]
	{\ensuremath{#1}}                                
\newcommand{\PRECVARXe}[2]
	{\ensuremath{\PRECVARX{#1} \langle #2 \rangle}}  

\newcommand{\PRECx}[2]
	{\ensuremath{\MUREC{#1} (#2)}}
\newcommand{\PRECtx}[2]
	{\ensuremath{\MUREC{#1} (#2)}}      
\newcommand{\PRECVARte}[2]
	{\GRECVARte{#1}{#2}}

%
\newcommand{\PNEWKW}{\KWORD{new}}
\newcommand{\PREGKW}{\KWORD{reg}}
\newcommand{\PINKW}{\KWORD{in}}
\newcommand{\PNEWs}[4]
	{\ensuremath{\PNEWKW\, (\AT{#1}{#2}, \ROLE{#3}) \,\PINKW\, #4}}  
\newcommand{\PNEWa}[4]
	{\ensuremath{\PREGKW\, \AT{#1}{#2}[\ROLE{#3}] \,\PINKW\, #4}}
\newcommand{\PNEWp}[4]
	{\ensuremath{\PNEWKW\, \AT{#1}{#2} \,\KWORD{with}\, [#3] \,\PINKW\, #4}}

\newcommand{\PJOIN}[2]
	{\ensuremath{\KWORD{join}\, #1[#2]}}  

%
\newcommand{\PLOCK}
	{\ensuremath{\blacktriangledown \,}}
\newcommand{\PUNLOCK}
	{\ensuremath{\blacktriangle}}
\newcommand{\PGET}[2]
	{\ensuremath{#1 \OPASSIGN get(\STATEVAR{#2})}}  
\newcommand{\PPUT}[2]
	{\ensuremath{put(#1, \STATEVAR{#2})}}           
\newcommand{\PLRECV}[3]
	{#1 [\ROLE{#2}, \ROLE{#3}] \, ? \blacktriangledown}

%
\newcommand{\PSNET}[1]
	{\ensuremath{#1}}     
\newcommand{\PSNETi}[2]
	{\ensuremath{#1_{#2}}}
\newcommand{\PICHAN}[2]
	{\ensuremath{\mathtt{I}(#1 [\ROLE{#2}])}}
\newcommand{\POCHAN}[2]
	{\ensuremath{\mathtt{O}(#1 [\ROLE{#2}])}}
\newcommand{\PNETQUEUE}[2]
	{\ensuremath{\langle #1 ; #2 \rangle}}

\newcommand{\RAYCOMMENT}[1]{~\\ \textbf{RAY:} #1}

\newcommand{\capabilities}{\mathtt{c}}
\newcommand{\GQueue}{h}
\newcommand{\monitorSet}{D}
\newcommand{\inTop}{\text{ in }}
\newcommand{\scribble}{Scribble}
\newcommand{\java}{{{\sc Java}}}
\newcommand{\eval}{\downarrow}
\newcommand{\ocaml}{{{\sc Ocaml}}}
\newcommand{\mN}{\mathsf{N}}

    \newcommand{\com}[2]{\par
      \fcolorbox{red}{yellow}{\parbox{\linewidth}{ 
            \color{gray}
            \begin{description}
            \item[{\color{blue} #2:}]{\sf #1}
            \end{description}}}
    }

\spnewtheorem{DEF}[theorem]{Definition}{\bfseries}{\rmfamily}
\spnewtheorem{REM}[theorem]{Remark}{\bfseries}{\rmfamily}
\spnewtheorem{PRO}[theorem]{Proposition}{\bfseries}{\rmfamily}
\spnewtheorem{CON}[theorem]{Convention}{\bfseries}{\rmfamily}
\spnewtheorem{LEM}[theorem]{Lemma}{\bfseries}{\itshape}
\spnewtheorem{THM}[theorem]{Theorem}{\bfseries}{\itshape}
\spnewtheorem{COR}[theorem]{Corollary}{\bfseries}{\itshape}
\spnewtheorem{EX}[theorem]{Example}{\bfseries}{\rmfamily}

\newcommand{\assleft}{\llbracket}
\newcommand{\assright}{\rrbracket}
\newcommand{\INTp}[1]{\II{#1}}
\newcommand{\OUTp}[1]{\OO{#1}}
\newcommand{\PENDINV}[4]{\{\AT{#1}{#2}[#3] \}_{#4}}

\newcommand{\NI}{\noindent}
\newcommand{\CD}{\!\cdot\!}
\newcommand{\CAL}[1]{\mathcal{#1}}
\newcommand{\OL}[1]{\overline{#1}}
\newcommand{\DIFF}{\backslash}
\newcommand{\Dropp}[2]{#1-#2}
\newcommand{\ifthenelse}{}
\newcommand{\VEC}{\tilde}
\newcommand{\VECw}{\widetilde}

\newcommand{\ENCan}[1]{\langle #1 \rangle}
\newcommand{\ENCda}[1]{\langle\!\langle #1 \rangle\!\rangle}
\newcommand{\ENCdp}[1]{(\!( #1 )\!)}

\newcommand{\ASET}[1]{\{#1\}}
\newcommand{\PAR}{\mathrel{\mid}}
\newcommand{\AT}[2]{#1\! : \! #2}

\newcommand{\OP}[2]{#1\ENCan{#2}}
\newcommand{\OT}[2]{#1(#2)} 
\newcommand{\MSG}[4]{\ENCan{#1, #2, \OP{#3}{#4}}}
\newcommand{\MSGT}[4]{\ENCan{#1, #2, \OT{#3}{#4}}}

\newcommand{\LT}{T} 

\newcommand{\GAbody}{\mathcal{H}}  
\newcommand{\GAssert}{G}  
\newcommand{\GA}{\GAssert}
\newcommand{\LAssert}{T}  
\newcommand{\LA}{\LAssert}
\newcommand{\LAbody}{\mathcal{U}}
\newcommand{\sort}{(\LAssert[\player])}
\newcommand{\MT}{\mathnormal{mv}} 
\newcommand{\AssertEnv}{\Gamma^A} 
\newcommand{\REFINES}{\Supset}
\newcommand{\CATCH}{~\mathtt{catch}~}
\newcommand{\CATCHAT}[2]{\CATCH\mathtt{at}~#1~\mathtt{to}~#2}

\newcommand{\mode}[1]{\keyword{m}(#1)}
\newcommand{\Imode}{\mathtt{I}}
\newcommand{\Omode}{\mathtt{O}}
\newcommand{\IOmode}{\mathtt{IO}}
\newcommand{\imode}{\mathsf{}}
\newcommand{\omode}{\mathsf{}}

\newcommand{\rro}{\ROLE{r}_1}
\newcommand{\rrt}{\ROLE{r}_2}
\newcommand{\rr}{\ROLE{r}}
\newcommand{\Buy}{\ROLE{B}}
\newcommand{\Sell}{\ROLE{S}}
\newcommand{\Agency}{\ROLE{A}}
\newcommand{\DB}{\ROLE{DB}}
\newcommand{\pr}{\alpha}
\newcommand{\prb}{\beta}

\newcommand{\brkin}{|}

\newcommand{\U}{\mode{\GA[\p]}}
\newcommand{\UI}{(\LA[\p])^\Imode}
\newcommand{\UO}{(\LA[\p])^\Omode}
\newcommand{\UIO}{(\GA[\p])^\IOmode}
\newcommand{\UIOG}{\GA[\p]}
\newcommand{\UIA}{\Imode(\assleft A\assright\GA[\p])}
\newcommand{\UA}{\mode{\assleft A\assright\GA[\p]}}
\newcommand{\UOA}{\Omode(\assleft A\assright\GA[\p])}

\newcommand{\VAR}[1]{var(#1)} 
\newcommand{\IVAR}[1]{\mathtt{fv}(#1)}  

\newcommand{\GInter}[2]{#1 \rightarrow #2}
\newcommand{\GForm}[7]{\GInter{#1}{#2}: \ASET{#3 (#4: #5) #6. #7}}
\newcommand{\NGForm}[6]{\GInter{#1}{#2}: \lbrace #3 (#4: #5) #6}
\newcommand{\LFormOut}[6]{#1! \ASET{#2 (#3: #4) #5. #6}}
\newcommand{\LFormIn}[6]{#1? \ASET{#2 (#3: #4) #5. #6}}
\newcommand{\LFormDaggerOne}[6]{#1\dagger_1 \ASET{#2 (#3: #4) #5. #6}}
\newcommand{\LFormDaggerTwo}[6]{#1\dagger_2 \ASET{#2 (#3: #4) #5. #6}}

\newcommand{\LFormOutMarked}[6]{\underline{#1 !} \ASET{#2 (#3: #4) #5. #6}}
\newcommand{\LFormInMarked}[6]{\underline{#1 ?} \ASET{#2 (#3: #4) #5. #6}}

\newcommand{\GInterMarked}[2]{\underline{#1 \rightarrow #2}}
\newcommand{\GFormMarked}[7]{\GInterMarked{#1}{#2}:\ASET{#3 (#4: #5) #6. #7}}

\newcommand{\Rec}[5]{\mu #1 ( #2 ) \ASET{#3} \ENCan{#4} . #5}
\newcommand{\RecDef}[2]{#1 \ENCan{#2}}
\newcommand{\RecDefV}[3]{#1 (#2) \ENCan{#3}}
\newcommand{\GSat}{\mathit{GSat}} 

\newcommand{\typing}[3]{#1\proves #2 \triangleright #3}
\newcommand{\MAssert}[4]{\ENCan{#1, #2, \OT{#3}{#4}}}

\newcommand{\s}{k} 
\newcommand{\sv}{y} 
\newcommand{\sn}{s} 

\newcommand{\aname}{u} 
\newcommand{\av}{x} 
\newcommand{\af}{a} 

\newcommand{\RecT}{\keyword{t}} 

\newcommand{\participant}[1]{\ensuremath{\mathtt{#1}}}
\newcommand{\q}{\ensuremath{\participant{q}}}
\newcommand{\p}{\ensuremath{\participant{p}}}
\newcommand{\player}{\participant{p}}
\newcommand{\ply}{\player} 

\newcommand{\PInter}[2]{#1 [#2]}
\newcommand{\BRANCHcp}[7]{\PInter{#1}{#2,#3}?\{ #4(#5).#6 \}_{#7}}
\newcommand{\BRANCHsig}[6]{\PInter{#1}{#2,#3}? #4(#5).#6}
\newcommand{\Branch}[6]{\{ #1 (#2\! :\!  #3)\{#4\}.#5\}_{#6}}

\newcommand{\SELECTcp}[6]{\PInter{#1}{#2,#3}! #4\ENCan{#5}; {#6}}
\newcommand{\SELECTdef}[5]{\PInter{#1}{#2,#3}! #4\ENCan{#5}}

\newcommand{\AOUTPUT}[4]{\overline{#1}\ENCan{#2[#3]:#4}} 
\newcommand{\AINPUT}[5]{{#1} (\AT{#2[#3]}{#4}). #5} 
\newcommand{\AINPUTT}[4]{{#1} (\AT{#2[#3]}{#4})} 
\newcommand{\ABOUTPUT}[2]{\overline{#1}(#2)}
\newcommand{\ABINPUT}[2]{{#1}(#2)}

\newcommand{\Qin}[2]{#1^\imode\!\!:\!#2}
\newcommand{\Qout}[2]{#1^\omode\!\!:\!#2}
\newcommand{\Qext}[2]{#1\!:\!#2}

\newcommand{\IF}[1]{{\text{if}}\ #1}
\newcommand{\IFTHENELSE}[3]{\text{if}\ #1\ \text{then}\ #2\ \text{else}\ #3}

\newcommand{\new}[3]{\keyword{new}\: \AT{#1}{#2}\ \keyword{in}\ #3}

\newcommand{\newandjoin}{\keyword{new}\ s : \{a_i : \LA_i[\rr_i]\}_{i\in I} \ \keyword{in}\ P}

\newcommand{\SJOIN}[3]{\mathsf{join}\ #1 [#2] ;#3}
\newcommand{\SLEAVE}[3]{\mathsf{leave}\ #1 [#2] ;#3}
\newcommand{\JOIN}[4]{#1 [#2]:#3 . #4}
\newcommand{\JOINP}[3]{\mathsf{join}\ #1\ \mathsf{as}\ #2\ \mathsf{in}\ #3}
\newcommand{\recur}[6]{(\mu #1 (#2\ \ #3).#4)\ENCan{#5, #6}}
\newcommand{\recurDef}[3]{#1 (#2\ \ #3)}
\newcommand{\recurIn}[2]{\ENCan{#1\ \ #2} }

\newcommand{\Nus}[2]{(\nu #1: #2)}
\newcommand{\Nua}[2]{(\nu #1: #2)}

\newcommand{\INACT}{\mathbf{0}}
\newcommand{\inact}{\mathbf{0}}

\newcommand{\proj}{\pmb{\pmb{\upharpoonright}}}
\newcommand{\Proj}[2]{#1 \upharpoonright {#2}}
\newcommand{\pmt}{\kw{pm}}
\newcommand{\PMForm}[3]{ #1 ! #2 (#3)}
\newcommand{\SelfType}[2]{\keyword{st}(#1, #2)}

\newcommand{\LInter}[2]{#1 [#2]}

\newcommand{\AOUT}[4]{\overline{#1}\ENCan{#2[#3]:#4}}
\newcommand{\AIN}[4]{#1\ENCan{#2[#3]:#4}}

\newcommand{\SOUTPUT}[5]{#1[#2,#3]!\OP{#4}{#5}} 
\newcommand{\SINPUT}[5]{#1[#2,#3]?\OP{#4}{#5}} 

\newcommand{\newl}[2]{\keyword{new}\ #1\ :\ #2}

\newcommand{\TRANS}[1]{\xrightarrow{#1}}
\newcommand{\TRANSS}[1]{{\xrightarrow{\raisebox{-.3ex}[0pt][0pt]{\scriptsize $#1$}}}}
\newcommand{\TAUTRANS}[1]{\stackrel{#1}{\Longrightarrow}}
\newcommand{\TAUTRANSh}[1]{\stackrel{\hat{#1}}{\Longrightarrow}}

\newcommand{\CHAIN}[1]{ (#1)^{\keyword{c}}}
\newcommand{\LCHAIN}[2]{ (#1)^{\keyword{c}_{#2}}}
\newcommand{\MCHAIN}[1]{ (#1)^{\keyword{cm}}}
\newcommand{\MLCHAIN}[2]{ (#1)^{\keyword{cm}_{#2}}}
\newcommand{\INCHAIN}[1]{ (#1)^{\keyword{ci}}}
\newcommand{\INLCHAIN}[2]{ (#1)^{\keyword{ci}_{#2}}}

\newcommand{\VECL}{\mathfrak{L}} 

\newcommand{\Prod}[1]{\prod_{#1}}

\newcommand{\GW}{{\mathbb{G}}}
\newcommand{\ACTIVE}[1]{#1} 
\newcommand{\NormalACT}[1]{#1} 
\newcommand{\INACTIVE}[1]{#1} 
\newcommand{\INFER}[2]{\frac{\displaystyle{#1}%
\vspace{2mm}%
}{
\vspace{2mm}%
\displaystyle{#2}
}}

\newenvironment{CASES}
{\left\{%
 \begin{array}{lc}}%
{\end{array}%
 \right.}

\newcommand{\CONFORMS}[2]{#1 \models #2}
\newcommand{\GTRANS}[1]{\TRANS{#1}_\text{g}}
\newcommand{\REL}[1]{\mathcal{#1}}
\newcommand{\RED}{\longrightarrow}

\newcommand{\keyword}[1]{\textsf{\upshape #1}}
\newcommand{\kw}{\keyword} 
\newcommand{\defk}{\keyword{def}}
\newcommand{\ink}{\keyword{in}}

\newcommand{\andk}{\keyword{and}}
\newcommand{\ork}{\keyword{or}}
\newcommand{\NOT}[1]{\;\;\;\not\!\!\!\!\!\!#1}

\newcommand{\truek}{\keyword{tt}}
\newcommand{\falsek}{\keyword{ff}}
\newcommand{\acceptk}{\keyword{accept}}
\newcommand{\requestk}{\keyword{request}}

\newcommand{\ifk}{\keyword{if}}
\newcommand{\thenk}{\keyword{then}}
\newcommand{\elsek}{\keyword{else}}
\renewcommand{\ifthenelse}[3]{\ifk\ #1\ \thenk\ #2\ \elsek\ #3}
\newcommand{\kend}{\keyword{end}}

\newcommand{\ADD}{\mathfrak{+}} 
\newcommand{\AND}{\wedge}
\newcommand{\ANDw}{\,\wedge\,}
\newcommand{\OR}{\vee}
\newcommand{\ORw}{\,\vee\,}
\newcommand{\DELETE}{\mathfrak{-} } 
\newcommand{\Retrv}{\mathfrak{-}}

\newcommand{\fnk}{\keyword{fn}}

\newcommand{\trule}[1]{{\footnotesize{\ensuremath{\lfloor\text{\sc{#1}}\rfloor}}}}
\newcommand{\mrule}[1]{{\footnotesize{\ensuremath{[\text{\sc{#1}}]}}}}

\newcommand{\srule}[1]{{\footnotesize{\ensuremath{(\text{\sc{#1}})}}}}

\newcommand{\monrule}[1]{{\footnotesize{\ensuremath{\lceil{\text{\sc{#1}}}\rceil}}}}

\newcommand{\lprule}[1]{{\footnotesize{\ensuremath{\{\text{\sc{#1}}\}}}}}

\newcommand{\natk}{\kw{nat}}
\newcommand{\boolk}{\kw{bool}}
\newcommand{\intk}{\keyword{int}}
\newcommand{\stringk}{\keyword{string}}

\newcommand{\defink}[2]{\defk\ #1\ \ink\ #2}
\def\eqdef{\;\stackrel{\text{\scriptsize def}}{=}\;}
\def\DEFEQ{\eqdef}

\newcommand{\proves}{\vdash}
\newcommand{\has}{\triangleright}
\newcommand{\IFF}{\Leftrightarrow}
\newcommand{\NULL}{\varepsilon}

\newcommand{\SB}{\sim}
\newcommand{\WB}{\approx}
\newcommand{\LITEQ}{\equiv}              
\newcommand{\Cong}{\equiv} 
\newcommand{\SUBS}[2]{[{#1}/{#2}]}
\newcommand{\MSUBS}[2]{\{{#1}/{#2}\}}
\newcommand{\OFF}[2]{#1 / #2}


\newcommand{\sbj}[1]{\mathsf{sbj}(#1)}
\newcommand{\domain}[1]{\mathsf{dom}(#1)}
\newcommand{\fn}[1]{\mathsf{fn}(#1)}
\newcommand{\bn}[1]{\mathsf{bn}(#1)}

\newcommand{\role}[1]{#1}
\newcommand{\names}[1]{\mathsf{n}(#1)}
\newcommand{\n}[1]{\mathsf{n}(#1)} 
\newcommand{\num}[1]{\mathsf{num}(#1)}

\newcommand{\erase}[1]{\mathsf{erase}(#1)}
\newcommand{\dom}[1]{\mathsf{dom}(#1)}
\newcommand{\complete}[1]{\keyword{COM}(#1)}
\newcommand{\permutableTo}{\curvearrowright}
\newcommand{\permutableToOnce}{\curvearrowright^1}

\newcommand{\LP}{\mathcal{L}}
\newcommand{\RPP}{RPP}
\newcommand{\Location}{\keyword{L}}
\newcommand{\capability}[1]{\mathsf{capability}(#1)}
\newcommand{\CAPABILITY}[1]{\mathsf{cap}(#1)}
\newcommand{\IMAGINARY}{{imaginary}}
\newcommand{\INNETWORK}{\mbox{in the network}} 
\newcommand{\OUTNETWORK}{\mbox{out of the network}} 

\newcommand{\Adaptor}{\mathfrak{A}}
\newcommand{\ini}{\keyword{ini}}

\newcommand{\JOINl}[1]{\mathsf{join}(#1)} 
\newcommand{\JSESS}[1]{#1^\bullet}

\newcommand{\localErr}{\mathsf{localErr}}
\newcommand{\envErr}{\mathsf{envErr}}
\newcommand{\Err}{\mathsf{err}}

\newcommand{\err}{\mathsf{err}}

\newcommand{\EVE}{\mathcal{E}}
\newcommand{\EVEG}{\Theta}
\newcommand{\EVED}{\Delta}
\newcommand{\EveLTS}{\GTRANS}

\newcommand{\pick}[2]{\rightsquigarrow(#1, #2)}

\newcommand{\SBeve}{\sim_{\mathsf{eve}}}
\newcommand{\WBeve}{\approx_{\mathsf{eve}}}

\newcommand{\AOUTNT}[3]{\overline{#1}\ENCan{#2[#3]}}
\newcommand{\NAOUTPUT}[2]{\overline{#1} \ENCan{#2}}
\newcommand{\AOUTPUTNT}[4]{\overline{#1}\ENCan{#2[#3]}; #4} 
\newcommand{\unmark}[1]{\mathsf{unmark}(#1)}

\newcommand{\GAmarked}{\underline{\GA}}  
\newcommand{\LAmarked}{\underline{\LA}}  

\newcommand{\TZ}[1]{#1}
\newcommand{\TZC}[1]{{#1}}

\newcommand{\JOINS}[3]{\kw{join}\ #1\ \kw{as}\ #2:#3}
\newcommand{\MCon}{{C}}

\newcommand{\GF}{\textbf{G}}

\newcommand{\pt}{\kw{pt}}

\newcommand{\namedP}[3]{[#2]_{#3}\mid #1}
\newcommand{\formalM}{\mon{\Gamma; \Delta}@\alpha}
\newcommand{\NP}[2]{[#1]_{#2}}

\newcommand{\namedPmo}[2]{[#1]_{#2}} 

\newcommand{\monitored}[3]{#1[{\mathcal{M}}(#2)\proves #3]}

\newcommand{\field}{\jmath}
\newcommand{\Leffect}[2]{\ENCan{#1\ , #2}}
\newcommand{\Reffect}[2]{\ASET{#1\ , #2}}
\newcommand{\LeffectRole}[3]{\ENCan{#1^{#3}\ , #2^{#3}}}
\newcommand{\ReffectRole}[3]{\ASET{#1^{#3}\ , #2^{#3}}}

\newcommand{\sumtpB}[9]{
#1 \! \rightarrow \! #2: \{ #3 \! (#4)
\! \LEFF{#5} \! \REFF{#7}.#9\}
}

\newcommand{\sumtpBcom}[8]
{
  #1 \! \rightarrow \! #2:
  \{ #3(#4)
  \REFF{{#5}}
  }

\newcommand{\lsumOut}[6]{
#1! \{ #2(#3) \REFF{#4}.#6 \}
}

\newcommand{\lsumIn}[6]{
#1? \{ #2(#3) \REFF{#4}.#6 \}}

\newcommand{\rectp}[5]{(\mu #1(#2)\ASET{#3}.#4)\ENCan{#5}}
\newcommand{\insttp}[2]{#1\ENCan{#2}}

\newcommand{\newname}[1]{\pmb{\nu}\, #1}

\newcommand{\prohibited}[2]{\keyword{prohibitedUnder}(#1, #2)}

\newcommand{\allowed}[2]{\keyword{allowed}(#1, #2)}
\newcommand{\after}[2]{\keyword{after}(#1, #2)}

\newcommand{\joinm}[2]{\mathit{join}\ENCan{#1[#2]}}
\newcommand{\jointype}[3]{\mathit{join}\ENCan{#1[#2]:#3}}
\newcommand{\leavem}[2]{\mathit{leave}\ENCan{#1[#2]}}
\newcommand{\leavetype}[3]{\mathit{leave}\ENCan{#1[#2]:#3}}
\newcommand{\newm}[1]{\mathit{new}\ENCan{#1}}

\newcommand{\newmNPD}[4]{\mathit{new}\ENCan{#1:#2[#3],#4}}

\newcommand{\Message}{M}

\newcommand{\newNP}[4]%
{\mathsf{new}\ #1 \ \mathsf{with} \ #2[#3]\ \mathsf{in} \ #4}

\newcommand{\newNPmo}[3]%
{\mathsf{new}\ #1 \ \mathsf{with} \ #2\ \mathsf{in} \ #3}

\newcommand{\newNPD}[5]%
{\mathsf{new}\ #1 \ \mathsf{with} \ #2[#3], #5\ \mathsf{in} \ #4}

\newcommand{\interfacedN}[2]{#1[#2]}

\newcommand{\AddressInterface}{\CAL{A}}
\newcommand{\AI}{\AddressInterface}
\newcommand{\BI}{\CAL{B}}
\newcommand{\Monitor}{\CAL{M}}

\newcommand{\MonitorG}{\mon{\Gamma}}

\newcommand{\namedM}[2]{{#1}}

\newcommand{\WFN}[2]{\proves \interfacedN{#1}{#2}\,:\,\diamond}
\newcommand{\ADR}[1]{\text{adr}(#1)}

\newcommand{\InputAct}[1]{\Downarrow #1}
\newcommand{\OutputAct}[1]{\Uparrow #1}
\newcommand{\TauAct}{\mathbf{\tau}}
\newcommand{\tauact}{\mathbf{\tau}}

\newcommand{\obj}[1]{\mathsf{obj}(#1)}

\newcommand{\NEWP}[2]{\mathsf{new}\ #1\ \mathsf{in}\ #2}

\newcommand{\newP}[2]{\mathsf{new}\ #1\ \mathsf{in}\ #2}

\newcommand{\destination}[1]{\mathsf{dest}(#1)}
\newcommand{\source}[1]{\mathsf{src}(#1)}
\newcommand{\Incoming}{\mathsf{incoming}}
\newcommand{\Outgoing}{\mathsf{outgoing}}

\newcommand{\INW}[2]{#1\proves #2}

\newcommand{\MAP}[1]{[\![#1]\!]}
\newcommand{\MAPenc}[1]{\ENCan{\!\ENCan{#1}\!}}

\newcommand{\monLeft}{\langle\!\langle}
\newcommand{\monRight}{\rangle \!\rangle}
\newcommand{\mon}[1]{\monLeft #1 \monRight}
\newcommand{\unmon}[1]{\mathsf{unmon}(#1)}


\newcommand{\getBI}[1]{\mathsf{int}(#1)}

\newcommand{\COMMENT}[1]{}
\newcommand{\PREDICATE}[1]{{\cal{#1}}}
\newcommand{\EFFECT}[1]{{\ENCan{#1}}}

\newcommand{\midW}{\,\mid\,}
\newcommand{\midWW}{\ \mid\ }

\newcommand{\coherent}{\asymp}

\newcommand{\GQ}[2]{\ENCan{#1\ ;\ #2}}

\newcommand{\smsg}[5]{#1\ENCan{#2,#3,#4\ENCan{#5}}}

\newcommand{\OUTPUT}[3]{\OL{#1}\ENCan{#2};#3}
\newcommand{\INPUT}[3]{#1(#2).#3}

\newcommand{\hastype}{\triangleright}

\newcommand{\effected}[2]{#1\,\mathtt{after}\,#2}

\newcommand{\DSP}{{\sf DSP}}

\newcommand{\PRG}{{P}}
\newcommand{\PRGQ}{{Q}}
\newcommand{\sN}{M}
\newcommand{\dN}{N}

\newcommand{\PRT}{P}

\newcommand{\PRTone}{P_{rt}}

\newcommand{\RInfo}{r}  

\newcommand{\DECLARE}[1]{(\!(#1)\!)}

\newcommand{\ONW}[2]{{#1}} 

\newcommand{\ENV}{\OL{N}}

\newcommand{\IMPLIES}{\Longrightarrow}

\newcommand{\NWSAT}[3]{#1 \models #2 : #3}
\newcommand{\NWSATa}[3]{#1 \models' #2 : #3}
\newcommand{\TRIPLE}[3]{\ENCan{#1, #2, #3}}

\newcommand{\goodP}[3]{#1\, :_#2\, #3}
\newcommand{\goodNW}[2]{#1\, :\, #2}
\newcommand{\goodNWunder}[3]{#1 \,\triangleright\, \goodNW{#2}{#3}}

\newcommand{\NWgood}[2]{#1 \,\triangleright\, #2}
\newcommand{\NLNWgood}[2]{#1 \,:\, #2}
\newcommand{\Ocompose}[2]{#1 \uplus #2}
\newcommand{\Icompose}[2]{#1 , #2}

\newcommand{\principals}[1]{\mathcal{P}(#1)}

\newcommand{\AEnv}{\mathcal{C}}
\newcommand{\InvExtract}[1]{\mathtt{inv}(#1)}
\newcommand{\AllExtract}[1]{\mathtt{inv}(#1)}
\newcommand{\offsession}[3]{#1[\ROLE{#2}]}
\newcommand{\onsession}[3]{#1[\ROLE{#2}]}
\newcommand{\InvNote}[1]{\tiny\blacksquare_{#1}}
\newcommand{\InvNotes}[1]{\tiny\widetilde{\blacksquare}_{#1}}
\newcommand{\InvA}{\textbf{A}}
\newcommand{\newNPmoPR}[3]%
{\mathsf{new}\ #1 \ \mathsf{with} \ [#2]_{\sigma'}\ \mathsf{in} \ #3}
\newcommand{\SEnv}{\mathcal{S}}
\newcommand{\state}[1]{S^{#1}}
\newcommand{\lstate}[1]{\locked{S}^{#1}}
\newcommand{\ustate}[1]{\unlocked{S}^{#1}}
\newcommand{\statep}[1]{S'^{#1}}
\newcommand{\lstatep}[1]{\locked{S'}^{#1}}
\newcommand{\ustatep}[1]{\unlocked{S'}^{#1}}

\newcommand{\OLtrule}[1]
{{\footnotesize{\ensuremath{\lfloor\OL{\text{\sc{#1}}\rfloor}}}}}

\newcommand{\LINKED}{{dynamic}}
\newcommand{\UNLINKED}{{static}}

\newcommand{\ROLES}[1]{{\mathsf{roles}(#1)}}

\newcommand{\OSPEC}{{\Theta^{\mathsf{ex}}}}
\newcommand{\ISPEC}{{\Theta^{\mathsf{in}}}}

\newcommand{\OSPECi}[1]{{\Theta^{\mathsf{ex}}_#1}}
\newcommand{\ISPECi}[1]{{\Theta^{\mathsf{in}}_#1}}

\newcommand{\NWsat}[3]{#1 \triangleright\, #2 \,:\, #3}

\newcommand{\EC}[1]{{\mathcal E}(#1)}
\newcommand{\EConly}{{\mathcal E}}

\newcommand{\NEC}[1]{{\mathcal N}(#1)}
\newcommand{\NEConly}{{\mathcal N}}

\newcommand{\LockP}{\mathcal{P}}
\newcommand{\LockPOut}{\LockP^\uparrow}
\newcommand{\LockPIn}{\LockP^\downarrow}
\newcommand{\Lock}{\blacktriangledown}
\newcommand{\Unlock}{\blacktriangle}
\newcommand{\Locking}[1]{\Lock #1 \Unlock}
\newcommand{\Evaluate}{\square}
\newcommand{\Read}{\boxdot}
\newcommand{\Write}{\boxtimes}

\newcommand{\GET}{\VEC{x}=get(\VEC{\field});}
\newcommand{\PUT}{put(\VEC{e}, \VEC{\field});}
\newcommand{\PUTUNLOCK}{put(\VEC{e}, \VEC{\field})\Unlock}
\newcommand{\UNLOCK}{\Unlock}

\newcommand{\GETxi}[2]{{#1}=get(#2)}
\newcommand{\PUTxi}[2]{put(#1, #2)}

\newcommand{\Locked}[1]{\underline{#1}}
\newcommand{\LockedInput}[6]
{#1[#2, #3]  ?\Lock \{ #4(#5). #6\}}

\newcommand{\ri}[1]{\mathsf{route}(#1)}
\newcommand{\control}{\mathsf{control}} 
\newcommand{\addinfo}{\varrho} 
\newcommand{\newsLabel}[4]{\keyword{new}(s) \{a_i: \LA_i[\rr_i]\} }
\newcommand{\newaLabel}[3]{\keyword{reg}~a :{#2}[#3]}
\newcommand{\newpLabel}[3]{\nu \alpha :{#2}[#3]} 
\newcommand{\pendingInvitations}[4]{\{a_i\mapsto #1[#2]:#3\}}

\newcommand{\Scribble}{Scribble}
\newcommand{\PARA}[1]{\paragraph*{{\bf #1.}}}

\newcommand{\Ser}{\texttt{S}}
\newcommand{\Cli}{\texttt{C}}
\newcommand{\Agent}{\texttt{A}}
\newcommand{\FON}[1]{{\mathtt{fobj}}(#1)} 

\newcommand{\nEC}[1]{{#1}}

\newcommand{\dual}{\mathit{dual}}





\newcommand{\M}{\mathsf{M}}
\newcommand{\Mp}{\mathsf{M}'}

\newcommand{\monitor}{\M}

\newcommand{\A}{\mathcal{M}^{\circ}}
\newcommand{\Ap}{\mathcal{M}_2^{\circ}}


\newcommand{\Moff}[1]{\monitor^{\circ}[#1]}
\newcommand{\Moffone}[1]{\monitor_1^{\circ}[#1]}
\newcommand{\Mofftwo}[1]{\monitor_2^{\circ}[#1]}
\newcommand{\Moffp}[1]{\monitor_2^{\circ}[#1]}
\newcommand{\Mon}[1]{\monitor[#1]}
\newcommand{\Monp}[1]{\monitor'[#1]}
\newcommand{\Monone}[1]{\monitor_1[#1]}
\newcommand{\Montwo}[1]{\monitor_2[#1]}

\newcommand{\ptp}[1]{{\participant{#1}}}

\newcommand{\pf}{\__{T}}

\newcommand{\Recursion}[4]{\mu \keyword{t}.#3}
\newcommand{\RecursionT}[3]{\mu \keyword{t}\ENCan{#1}(#2).#3}
\newcommand{\RecursionP}[3]{\RecursionPAbs{#1}{#2}}
\newcommand{\RecursionPAbs}[2]{\mu X.#2}
\newcommand{\CallT}[1]{\keyword{t}}
\newcommand{\Call}[1]{\keyword{t}}
\newcommand{\CallP}[1]{X}
\newcommand{\typeconst}[1]{\mathsf{nat}}
\newcommand{\nat}{\typeconst{nat}}
\newcommand{\intTp}{\textsf{Int}}
\newcommand{\bool}{\textsf{Bool}}

\newcommand{\NTRANS}[1]{\stackrel{#1}{\not\longrightarrow}}
\newcommand{\const}[1]{\mathtt{#1}}
\newcommand{\ENTAILS}{\supset}
\newcommand{\MPSA}{{\it MPSA}}
\newcommand{\MPST}{{\it MPST}}

\newcommand{\MPSTs}{{\it MPST}s}
\newcommand{\MPSAs}{{\it MPSA}s}

\newcommand{\eveproves}{\proves_{\text{g}}}

\newif\ifny\nytrue
\nytrue
\newcommand{\NY}[1]
	{\ifny{\color{violet}{#1}}\else{#1}\fi}
\newcommand{\KH}[1]
	{\ifny{\color{red}{#1}}\else{#1}\fi}
\newcommand{\TZU}[1]
	{\ifny{\color{blue}{#1}}\else{#1}\fi}
\definecolor{dkgreen}{rgb}{0,0.5,0}
\newcommand{\MD}[1]
	{\ifny{\color{dkgreen}{#1}}\else{#1}\fi}
\newcommand{\RH}[1]
	{\ifny{\color{orange}{#1}}\else{#1}\fi}

\newcommand{\GC}{\mathcal{G}}
\newcommand{\LC}{\mathcal{T}}

\newcommand{\grmor}{\ \mathrel{\text{\large$\mid$}}\ }

\newcommand{\MPSAenv}{\envT{\Gamma}{\Delta}{D}}
\newcommand{\SPECenv}{\Theta}

\newcommand{\envTriple}[3]{#1;\, #2 #3}
\newcommand{\envT}[3]{\envTriple{#1}{#2}{#3}}
\newcommand{\envTT}[3]{\ENCan{\envTriple{#1}{#2}{#3}}}

\newcommand{\envQuintuple}[5]{#1;\, #2;\, #3;\,#4;\, #5}
\newcommand{\envQ}[5]{\envQuintuple{#1}{#2}{#3}{#4}{#5}}
\newcommand{\envQQ}[5]{\ENCan{\envTriple{#1}{#2}{#3}{#4}{#5}}}


\newcommand{\LNW}[4]{(\nu #1)(#2;\ \GQ{#3}{#4})}
\newcommand{\LNWn}[3]{(\nu \VEC{n})(#1;\ \GQ{#2}{#3})}

\newcommand{\ULNW}{\kw{N}}

\newcommand{\ULspec}{\kw{\Theta}}

\newcommand{\monitoredP}[3]{#1[#2]_{#3}}

\newcommand{\nubf}{{\pmb \nu}}

\newcommand{\MRED}{\rightarrow\!\!\!\!\!\rightarrow}

\newcommand{\AssEnv}{\ENCan{\Gamma;\Delta}}

\newcommand{\static}[1]{#1}
\newcommand{\dynamic}[1]{#1}

\newcommand{\sta}[1]{\static{#1}}
\newcommand{\dyn}[1]{\dynamic{#1}}

\newcommand{\codom}[1]{\mathsf{codom}(#1)}
\newcommand{\SHS}{Checkability}
\newcommand{\sHS}{checkability}
\newcommand{\STS}{TS}
\newcommand{\WTS}{WTS}

\newcommand{\TSWI}[3]{\STS(#1, #2, #3)}

\newcommand{\TSG}[2]{\WTS(#1)_{#2}}
\newcommand{\TSGI}[3]{\WTS(#1)_{#2}^{#3}}
\newcommand{\HSEnv}{{E}}
\newcommand{\intVar}[1]{\#(#1)}
\newcommand{\stVar}[1]{sv(#1)}
\newcommand{\stateVar}[1]{\mathtt{split_{stateVar}}(#1)}
\newcommand{\inv}[1]{\mathtt{inv}(#1)}
\newcommand{\svar}[1]{\mathtt{svar}(#1)}
\newcommand{\MYSTATEVAR}[2]{\kw{#2}}
\newcommand{\dec}{\mathbf{\mathtt{d}}}
\newcommand{\romain}[1]{$\mathbf{RD}$: #1 $\bigstar$}
\newcommand{\many}[1]{\overtilde{#1}}
\newcommand{\II}[1]{\mathtt{I}(#1)}
\newcommand{\OO}[1]{\mathtt{O}(#1)}

\newcommand{\SERVERROLE}{\texttt{S}}
\newcommand{\CLIENTROLE}{\texttt{C}}
\newcommand{\rumi}[2]{$\mathbf{RN}$:#1 \textcolor{gray}{#2} $\bigstar$}
\newcommand{\bparagraph}[1]{\noindent{\textbf{#1.}\ }}

\definecolor{dkblue}{rgb}{0,0.1,0.5}
\definecolor{dkgreen}{rgb}{0,0.4,0}
\definecolor{dkred}{rgb}{0.4,0,0}
\lstnewenvironment{PYTHONLISTING}%
{
\lstset{
  language=python,
  showstringspaces=false,
  formfeed=\newpage,
  tabsize=2,
  commentstyle=\color{dkgreen}\itshape,
  basicstyle=\ttfamily,
  morekeywords={models, lambda, forms, def, class}
  keywordstyle=\color{dkblue},
  emph={access,and,as,break,class,continue,def,del,elif,else,%
	except,exec,finally,for,from,global,if,import,in,is,%
	lambda,not,or,pass,print,raise,return,try,while,assert,with},
  emphstyle=\color{dkblue}\bfseries,
  basicstyle=\CODESTYLE\footnotesize,
  keywordstyle=\CODESTYLE\footnotesize
}
}
{
}

\newcommand{\rtsyntax}[1]{\colorbox{lightgray}{\ensuremath{#1}}}
\newcommand{\upd}{\mathit{update}}
\newcommand{\eq}{\ensuremath{\diameter}}
\newcommand{\kf}[1]{\textup{\textsf{#1}}\xspace}
\newcommand{\constf}[1]{\textup{\textsf{#1}}}
\newcommand{\srsimple}[3]{\ensuremath{\bar{#1}[#2](#3)}}
\newcommand{\sr}[4]{\ensuremath{\srsimple{#1}{#2}{#3}.#4}}
\newcommand{\uu}{\ensuremath{u}}
\newcommand{\Ia}{\ensuremath{a}}
\newcommand{\Ic}{\ensuremath{c}}
\newcommand{\Ias}{\ensuremath{\alpha}}
\newcommand{\Ib}{\ensuremath{b}}
\newcommand{\y}{\ensuremath{y}}
\newcommand{\PP}{\ensuremath{P}}
\newcommand{\Q}{\ensuremath{Q}}
\newcommand{\R}{\ensuremath{R}}
\newcommand{\h}{\ensuremath{h}}
\newcommand{\queue}{\ensuremath{\h}}
\newcommand{\DD}{\ensuremath{D}}
\newcommand{\sub}[2]{\ensuremath{\{#1/#2\}}}
\newcommand{\sasimple}[3]{\ensuremath{#1[#2](#3)}}
\newcommand{\sa}[4]{\ensuremath{\sasimple{#1}{#2}{#3}.#4}}
\newcommand{\redsym}{\ensuremath{\longrightarrow}}
\newcommand{\nm}[1]{{\texttt{\scriptsize #1.}}}
\newcommand{\ty}{\textbf{t}}
\newcommand{\End}{\kf{end}}
\newcommand{\pro}[2]{\ensuremath{#1\upharpoonright#2}}
\newcommand{\G}{\ensuremath{G}}
\newcommand{\Gv}[4]{\ensuremath{#1\to\pset:\langle#3\rangle.#4}}
\newcommand{\PBig}{\ensuremath{P}}
\newcommand{\red}[2]{\ensuremath{#1\redsym#2}}
\newcommand{\X}{\ensuremath{X}}
\newcommand{\x}{\ensuremath{x}}
\newcommand{\pset}{\ensuremath{\Pi}}
\newcommand{\sI}[1]{\ensuremath{\s_{#1}}}
\newcommand{\pI}[1]{\ensuremath{\PBig_{#1}}}
\newcommand{\kk} {\ensuremath{\kappa}}
\newcommand{\si}[2]{\ensuremath{#1[#2]}}
\newcommand{\ki}[2]{\ensuremath{#1(#2)}}
\newcommand{\valheap}[3]{\ensuremath{( #3,\pset,#1 )}}
\newcommand{\va}{\ensuremath{v}}
\newcommand{\sd}[4]{\ensuremath{#1!\langle\! \langle#3,#2\rangle \!\rangle;#4}}
\newcommand{\delheap}[3]{\ensuremath{(#3,{#2},#1)}}
\newcommand{\labheap}[3]{\ensuremath{( #3,\pset,#1 )}}
\newcommand{\ptilde}[1]{{\ensuremath{#1}}}
\newcommand{\at}[1]{\ensuremath{\ptilde{#1}}}
\newcommand{\proccalldots}[3]{\ensuremath{#1\langle\ptilde{#2},\ptilde{#3}\rangle}}
\newcommand{\Xsignature}{\ensuremath{\X(\at{x}, \at{y})}}
\newcommand{\Ddef}{\ensuremath{\Xsignature=\PP}}
\newcommand{\defX}{\ensuremath{\kf{def} \ \Ddef\ \kf{in}\ }}
\newcommand{\anglep}[2]{\ensuremath{\langle #1, #2\rangle}}
\newcommand{\set}[1]{\ensuremath{\{#1\}}}
\newcommand{\T}{\ensuremath{T}}
\newcommand{\UT}{\ensuremath{U}}
\newcommand{\oT}[2]{\ensuremath{\;!\langle #2,#1\rangle}}
\newcommand{\seltype}{\ensuremath{\oplus \langle \pset,\{l_i:\T_i\}_{i\in
I} \rangle }}
\newcommand{\Par}{\ensuremath{\ |\ }}
\newcommand{\pc}{\Par}
\newcommand{\qtail}[1]{\ensuremath{\qcomp{\queue}{#1}}}
\newcommand{\qhead}[1]{\ensuremath{\qcomp{#1}{\queue}}}
\newcommand{\mqueue}[2]{\ensuremath{#1 : #2}}
\newcommand{\qcomp}[2]{\ensuremath{#1 \cdot #2}}
\newcommand{\Th}{\ensuremath{\Theta}}
\newcommand{\dere}[2]{\ensuremath{\Th\,\vdash #1\; \blacktriangleright\; #2 \,;\,\M\,;\,\B}}

\newcommand{\der}[3]{\ensuremath{#1\vdash#2\triangleright#3}}
\newcommand{\Ga}{\ensuremath{\Gamma}}
\newcommand{\D}{\ensuremath{\Delta}}
\newcommand{\sid}[1]{\ensuremath{\textup{pn}(#1)}}
\newcommand{\de}[3]{\ensuremath{#1\vdash#2:#3}}
\newcommand{\ST}{\ensuremath{S}}
\newcommand{\pn}{\p}
\newcommand{\an}[1]{\ensuremath{\langle #1\rangle}}
\newcommand{\ins}{\ensuremath{:}}
\newcommand{\B}{\ensuremath{\mathcal{B}}}
\newcommand{\dereb}[3]{\ensuremath{\Th\,\vdash #1 \;\blacktriangleright\; #2 \,;\,\M\,;\,#3}}
\newcommand{\Or}{\ensuremath{\mathcal{R}}}
\newcommand{\nk}[1]{\ensuremath{\ell(#1)}}
\newcommand{\out}[4]{\ensuremath{#1!\langle \pset,#2\rangle;#4}}
\newcommand{\outp}[3]{\ensuremath{#1!\langle \pset,#2\rangle}}
\newcommand{\outs}[4]{\ensuremath{#1!\langle #3,#2\rangle;#4}}
\newcommand{\e}{\ensuremath{e}}
\newcommand{\adde}[2]{\ensuremath{#1\bar{\cup}\set{#2}}}
\newcommand{\cc}{\ensuremath{c}}
\newcommand{\prule}[1]{\{\text{\textup{\sc{#1}}}\}}
\newcommand{\pre}[2]{\ensuremath{\mathsf{pre}({#1},{#2})}}
\newcommand{\inp}[4]{\ensuremath{#1?( #3,#2);#4}}
\newcommand{\iT}[2]{\ensuremath{?( #2,#1 )}}
\newcommand{\rd}[4]{\ensuremath{#1?(\!(#3,#2)\!);#4}}

\newcommand{\GAs}{G} 

\lstset{ %
language=Java ,                
basicstyle=\footnotesize,       
backgroundcolor=\color{white},  
showspaces=false,               
showstringspaces=false,         
showtabs=false,                 
frame=single,                   
tabsize=2,                      
captionpos=b,                   
breaklines=true,                
breakatwhitespace=false,        
title=\lstname,                 
escapeinside={\%*}{*)},         
morekeywords={*,...}            
}

\title{Session Types Go Dynamic or\\  
How to Verify Your Python Conversations}
\author{
Rumyana Neykova
\institute{
Imperial College London,
United Kingdom \email{rumi.neykova10@imperial.ac.uk}}}
\def\titlerunning{Session Types Go Dynamic}
\def\authorrunning{R. Neykova}

\maketitle
\begin{abstract}
This paper presents the first implementation of session types in a dynamically-typed language - Python. Communication safety of the whole system is guaranteed at runtime by monitors that check the execution traces comply with an associated protocol. Protocols are written in Scribble, a choreography description language based on multiparty session types, with addition of logic formulas for more precise behaviour properties. The presented framework overcomes the limitations of previous works on the session types where all endpoints should be statically typed so that they do not permit interoperability with untyped participants. The advantages, expressiveness and performance of dynamic protocol checking are demonstrated through use case and benchmarks.
\end{abstract}
\section{Introduction}
\label{sec:intro}
The study of multiparty session types (MPST) has explored a type theory for distributed programs 
which can ensure, for any typable programs, a full guarantee of deadlock-freedom and communication safety (all processes conform to a globally agreed communication protocol) through static type checking. However, a static verification is not always feasible and dynamic approaches have several advantages. First, when access to the source code is restricted dynamic verification enables to detect and ensure the correctness of external untyped components. Second, constraints on the message payload are easier to check dynamically. Third, as shown in this paper, dynamic checking is less obstructive to the source code, because it does not require extensions of the host language as in the existing works on session types. 

\begin{wrapfigure}{r}{0.35\textwidth}
\centering
\includegraphics[scale=0.35]{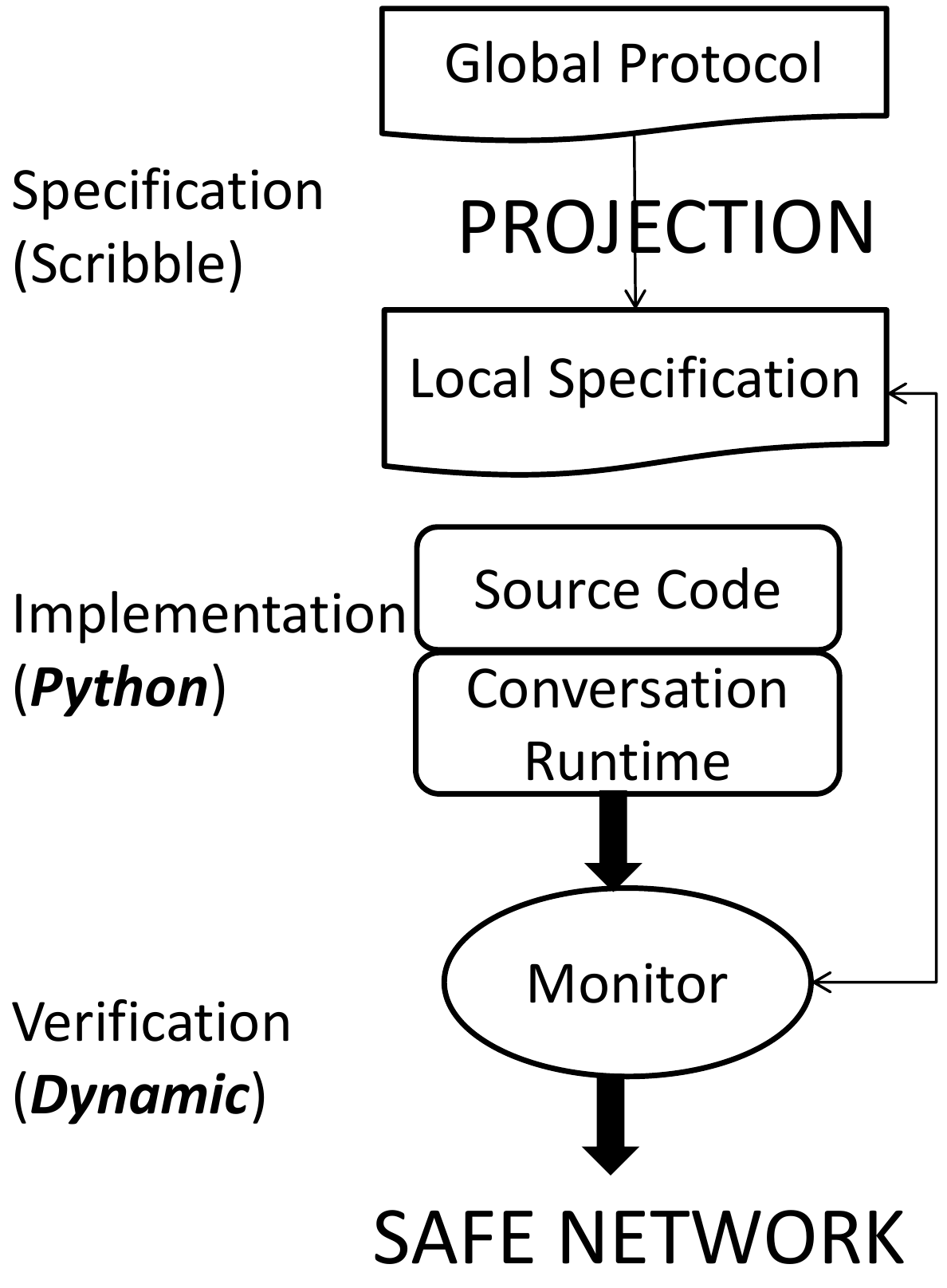}
\caption{Development methodology\label{fig:framework}}
\end{wrapfigure}
In this paper we present a toolchain for session-based programming (hereafter conversation programming) in Python that uses MPST-protocols to dynamically verify the communication safety of the running system. Conversation programming in Python resembles the standard development methodology for MPST-based frameworks (Fig.~\ref{fig:framework}). It starts by specifying the intended interactions (choreography) as a global protocol in the protocol description language Scribble \cite{Scribble}. Then Scribble local protocols are generated mechanically for each participant (role) defined in the protocol. After that  processes for each role are implemented  using  MPST operations exposed by Python conversation library. An external monitor is assigned to each endpoint.\\
During communication initiation the monitor retrieves the local protocol for its process and converts it to a finite state machine (FSM). The FSM continuously checks at runtime that each interaction (execution trace) is correct. If all participants comply to their protocols, the whole communication is guaranteed to be safe \cite{TGC11}. If participants do not comply, violations (such as deadlocks and communication mismatch) are detected and optionally ignored. 

The presented framework brings several non-trivial contributions to MPST works. First, Scribble is extended with logic assertions (constraints on the message payload). Second, implementing MPST in a dynamic language requires different code augmentation techniques. For that purpose, we have defined a minimal, but sufficient and extendable format for conversation message headers. Third, we show that using FSMs for MPST checking has reasonable overhead. The algorithm used to convert local session types to FSM is based on \cite{DY12}, however we have optimised it to avoid the state explosion for parallel sub-protocols and have extended it for the new Scribble constructs. 
Finally, the Python API is more flexible compared to other session types language extensions, because it supports different programming styles (event-driven and thread-based,  see Fig.~\ref{fig:api_implementation}). From the existing implementations only SJ \cite{event} features event-driven programming, but it has more strict typing rule.  
To the best of our knowledge, this is the first implementation of session types for decentralised monitoring. Our practical framework is inspired by the formal model of MPST runtime safety enforcement presented in \cite{TGC11,TC2013}. In the aforementioned works conformance to stipulated global protocols is guaranteed at runtime through local monitoring. 

The rest of the paper illustrates the key features of our conversation framework, the Python runtime and its API (\REF{sec:example}), it also gives overview of the monitoring tool, along with its benchmarks (\REF{sec:monitor}). \REF{sec:conclusion} discusses future work and concludes. The code for the runtime and the monitor tool and example applications are available from \cite{spy}. 

\section{Conversation Programming in Python}
\label{sec:example}
This section illustrates the stages of our framework and its implementation through a use case.
Step 1 and 2 illustrate the use case specification in Scribble, while Step 3  presents one of the main contributions of the paper -- a python API for conversation programming.  We present a use case obtained from our industrial partners Ocean Observatory Institute (OOI) \cite{OOI} (use case UC.R2.13 "Acquire Data From Instrument"). 
OOI aims to establish cyberinfrastructure for the delivery, management and
analysis of scientific data from a large network of ocean sensor. Their architecture relies on distributed run-time monitoring to regulate
the behaviour of third-party applications within the system. Part of the monitor tool presented in this paper is already integrated in their system as an internal monitor. \\

\begin{figure}[t]
\begin{minipage}[t]{0.5\textwidth}
{\lstset{numbers=left}
\begin{SJLISTING}
global protocol DataAquisition(role U, 
	role A, role I) {
  Request(string:info) from U to A; 
  Request(string:info) from A to I;	
  choice at I { 
	Support from I to A;
	rec Poll{  
	  Poll from A to I;
	  choice at I {
			@{size(data) <= 512}	    
			Raw(data) from I to A ;
			Formatted(data) from I to U;	
			Poll;
      } or { 
      	Stop from I to A;
        Stop from A to U;}} 
  } or { 
  	NotSupported from I to A;		
    Stop from A to I;	
    Stop from A to U;}}
\end{SJLISTING}}
\end{minipage}
\begin{minipage}[t]{0.5\textwidth}
{\lstset{numbers=left}
\begin{SJLISTING}
local protocol DataAquisition at A(role U, 
	role A, role I) {
  Request(string:info) from U; 
  Request(string:info) to I;	
  choice at I { 
	Support from I;
	rec Poll{  
	  Poll to I;
	  choice at I {
			@{size(data) <= 512}	    
			Raw(data) from I;	
			
			Poll;
      } or { 
        Stop from I;
        Stop to U;}} 
  } or { 
    NotSupported from I;		
    Stop to I;	
    Stop to U;}}
\end{SJLISTING}}
\end{minipage}
\vspace{-25pt}
\caption{Global Protocol (left) and Local Protocol for role A (right) \label{fig:protocol}}
\end{figure}

\bparagraph{Step 1: Global Protocol}
The Scribble global protocol for the use case is listed in Fig.~\ref{fig:protocol}.
Scribble describes interactions between session participants through message passing sequences, branches and recursion. Each message has a label (an operator) and a payload.
The first line declares the Data Acquisition protocol and three participant roles -- a User (U), an Agent service (A) and an Instrument (I). %
The overall scenario is as follows: U requests via A to start streaming a list of resources from I
(line 2--3). At Line 4 I makes a choice wether to continue the interaction or not. 
If I supports the requested  resource the communication continues and A starts to poll resources from I and streams them to U (line 6--15). Line 10 shows the new assertion construct and restricts I to send data packages that are less than 512MB. The presented assertion extension is inspired by \cite{BHTY10}. However, we do not stick to a predefined logic, but allow various policy languages to be incorporated inside an assertion construct. 

\bparagraph{Step 2: Global-to-local Protocol Projection}
Local protocols specify the communication behaviour for each conversation participant.
An example of a local protocol (the local protocol for role A is given in Fig.~\ref{fig:protocol}. A local protocol is essentially a view of the global protocol from the perspective of one participant role and as such it is mechanically projected from the global protocol. Projection basically
works by identifying the message exchanges where the participant is involved, and disregarding the rest, while preserving the overall interaction structure of the global protocol. The assertions are similarly preserved by projection where relevant. 
\label{sec:runtime}

\begin{wrapfigure}[13]{r}{0.4\textwidth}
\begin{PYTHONLISTING}
# session initiation bla 
create(protocol, inv_config.yml)
# accept an invitation
join(self, role, principal_name)
# send a msg
send(self, to_role, op, payload)
# receive a msg
recv(self, from_role) 
# receive asynchronously
recv_async(self, from_role, callback) 
# close the connection
stop() 
\end{PYTHONLISTING}
\vspace{-28pt}
\caption{Conversation API\label{fig:api-mapping}}
\vspace{-30pt}
\end{wrapfigure}

\bparagraph{Step 3: Process Implementation} 
Fig.~\ref{fig:api_implementation} illustrates the conversation API by presenting two alternative implementations in Python for the User process. Our Python conversation API offers a high level interface for safe conversation programming and maps basic session calculus primitives to
lower-level communication actions on a concrete transport (AMQP \cite{AMQP} in this case). The
implementation is built on top of Pika \cite{PIKA}, a widely used AMQP
client library for Python. Fig.~\ref{fig:api-mapping} lists the basic API methods.
In short, the API provides functionality for
(1) session initiation and joining and
(2) basic send/receive.
Each message embeds in its payload a conversation header. The header contains session information either for monitor initialisation (in case of \emph{invitation} messages), or  session checking (in case of \emph{in-session} messages).

\begin{figure}[t]
\begin{minipage}[t]{0.48\textwidth}
\begin{PYTHONLISTING}
class ClientApp(BaseApp):
	def start(self):
		c = Conversation.create('DataAquisition',  'config.yml')
		c.join('U', 'alice') 		

		resource_request = c.receive('U')
		c.send('I', resource_request)
		req_result = c.receive('I')

		if (req_result == SUPPORTED):
			c.send('I', 'Poll')
			op, data = c.receive('I')			
			while (op !=	'Stop'):		
				formatted_data = format(data)
				c.send('U', fomratted_data)
			c.send('U', stop)					
		else:
		    c.send('U, I', stop)
		    c.stop()
\end{PYTHONLISTING}
\end{minipage}
\begin{minipage}[t]{0.52\textwidth}
\begin{PYTHONLISTING}
class ClientApp(BaseApp):
	def start(self):
		c = Conversation.create('DataAquisition', 
						'config.yml')
		c.join('U', 'alice') 		
		c.receive_async('U', on_request_received)
	
	def on_request_received(self, conv, op, msg):
		if (op == SUPPORTED):
			conv.send('I', 'Poll')
			conv.receive_async('I', 'on_data_received')
		else: conv.send('I, U', 'Stop') 
	
	def on_data_received(self, conv, op, payload):
		if (operation !=	'Stop'):
			formatted_data = format(payload)			
			c.send('U', formatted_data)
		else:
		    conv.send('U', 'Stop')
		    conv.stop()
\end{PYTHONLISTING}
\end{minipage}
\vspace{-25pt}
\caption{Python standard (left) and event-driven (right) implementation of the User process \label{fig:api_implementation}}
\end{figure}

\textit{Conversation initiation} The \CODE{Conversation.create} method initiates a new conversation.
It creates a fresh conversation id and the required AMQP objects (principal exchange and queue), and sends an invitation message for each role specified in the protocol. Invitation mechanism is needed to map the role names to concrete addressable entities on the network (principals) and to propagate this mapping to all participants. Invitation header carries a conversation id, a role, a principal name (resolvable to a network address) and a name for a Scribble local specification file. In our example, the User starts a session and sends invitation to all other participants. 
Once the invitations are sent and accepted, a session is established and the intended message exchange can start.
An invitation for a role is accepted using the \CODE{Conversation.join} method. It establishes an AMQP connection and, if one does not exist, creates an invitation queue on which the invitee waits to receive an invitation. 

\textit{Conversation message passing}
The API provides standard send/receive primitives.  Send is
asynchronous, meaning that a basic send does not block on the
corresponding receive; however, the basic receive does block until the
complete message has been received.  An asynchronous receive
(\CODE{receive\_async}) is also provided to support event-driven usage
of the conversation API.  We have demonstrated two different implementations for the the User process (threaded and event-driven). Both versions require the same monitor for checking. The primitives for sending and receiving specify the name of the sender and receiver role respectively. 
The runtime resolves the role name to the actual network destination by coordinating with the in-memory conversation routing table created as a result of the conversation invitation. All messages are sent/received as a tuple of an operation and a payload. The API does not mandate how the operation field should be treated, allowing the runtime freedom to interpret the operation name various ways, e.g.\ as a plain message label, an RMI method name, etc. Syntactic sugar such as automatic dispatch on method calls based on the message operation is possible. 
More examples of programs using the API can be found in \cite{spy}.

\section{Dynamic Verification}
\label{sec:monitor}
\label{subsec:runtime}
\subsection{Monitoring Implementation}

To guarantee global safety our monitoring framework imposes {\em complete mediation}
of communications: no communication action should have an effect unless the message is mediated by the monitor. We use the AMQP's functions to reroute each outgoing/incoming message to its associated monitor. Routing is configured during session initialisation.
\begin{figure}[t]   
\centering
\includegraphics[scale=0.4]{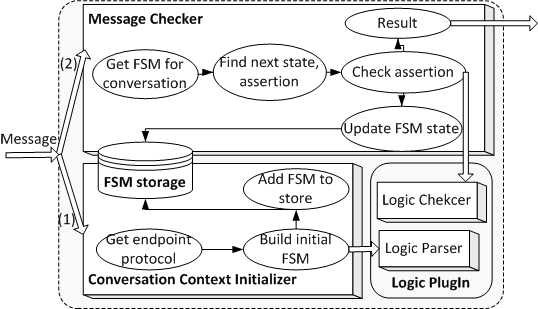}
\caption{Monitor components and workflow. The messages are processed depending on their type: \newline
 (1) Invitation Messages and (2) Conversation Messages.\label{fig:monitor:components}}
\end{figure}

Figure~\ref{fig:monitor:components} depicts the main components and internal
workflow of our prototype monitor. The lower part relates to
session initiation. The invitation message
carries (a reference to) the local type for the invitee and the session
id (global types can be exchanged if the monitor has the facility for
projection.) The monitor generates the FSM from the local type following
\cite{DY12}. Our implementation differs from \cite{DY12}
in the treatment of parallel sub-protocols (i.e.~unordered message
sequences). For efficiency, the monitor generates nested FSMs for each
session thread, avoiding the potential state explosion that comes from
constructing their product. FSM generation has therefore polynomial
time and space cost in the length of the local type. The (nested) FSM is
stored in a hash table with session id as the key. Due to MPST
well-formedness conditions (message label distinction), any nested FSM is uniquely
identifiable from any unordered message (i.e. session FSMs are deterministic).
Transition functions are similarly hashed, each entry having the shape:
$(\text{\em current$\underline{\ }$state},\ \text{\em transition})
\;\mapsto\;
(\text{\em next$\underline{\ }$state},\
\text{\em assertion},\ \text{\em var})$
 where $\text{\em transition}$ is a triple $(\text{\em label}, \text{\em sender}, \text{\em receiver})$, and \emph{var} is the variable binder
for the message payload.

The upper part of the Figure relates to in-session messages, which carry
the session id (matching an entry in the FSM hash table), sender and
receiver fields, and the message label and payload. This information allows
the monitor to retrieve the corresponding FSM (the message signature is
matched to the FSM's transition function). Any associated assertions are
evaluated by invoking an external logic engine; a monitor can be configured
to use various logic engines, for example, logic engines that support
the validation of assertions, automata-based specifications (such as security
automata), or state updates.
The current implementation uses a Python predicate
evaluator, which is sufficient for the example
protocol specifications that we have tested so far.





\subsection{Benchmarks}
\label{sec:benchmarks}
These benchmarks measure the communication
overhead introduced by our prototype monitor implementation. The
results show that the core FSM-related functionality of the monitor
adds little overhead in comparison to a dummy monitor that performs
plain message forwarding.

\bparagraph{Benchmark framework}
We measure the time to complete a
session between client and server endpoints connected to a single-broker AMQP network. Three benchmark cases are compared. The main case
(Monitor) is fully monitored,
i.e.\ FSM generation and message checking
are enabled for both the client and server.
The base case for comparison (Forwarder)
has the client and server in the same configuration, but with dummy
monitors that perform only message forwarding. For reference, the
final case (No Monitor) tests direct AMQP communication between the
server and client, i.e.\ messages are routed directly from an exchange
to their destination queues (no intermediate forwarding). Naturally,
forwarding-based
mediation incurs additional latencies; the
actual internal overhead of the monitor is
given by
the first two benchmark
cases.
This benchmark framework is applied to three scenarios:
\begin{enumerate}
\item Increasing \emph{session length} (number of messages), for protocol:
\vspace{0.5mm}

\centerline{$\mu \: X . \GInter{\mathtt{S}}{\mathtt{C}} \{ \mathtt{OK}() . \GInter{\mathtt{C}}{\mathtt{S}} \{ \mathtt{ACK}() . X \}, \mathtt{KO}() . \mathsf{end} \}$}

\noindent Session length is the number of times the recursion is repeated.
\item Increasing \emph{protocol size} (increasing number of parallel states). We repeatedly compose the base pattern to construct bigger protocols for nested FSM generation.
\vspace{0.5mm}

\centerline{$\GInter{\mathtt{S}}{\mathtt{C}} \{ \mathtt{OK}() . \mathsf{end} \} \;|\; \GInter{\mathtt{C}}{\mathtt{S}} \{ \mathtt{ACK}() . \mathsf{end} \}$}
\item Increasing \emph{payload size} (message size), using protocol from (1).
\end{enumerate}
\begin{figure}[t]
\centering
    \includegraphics[width=27em]{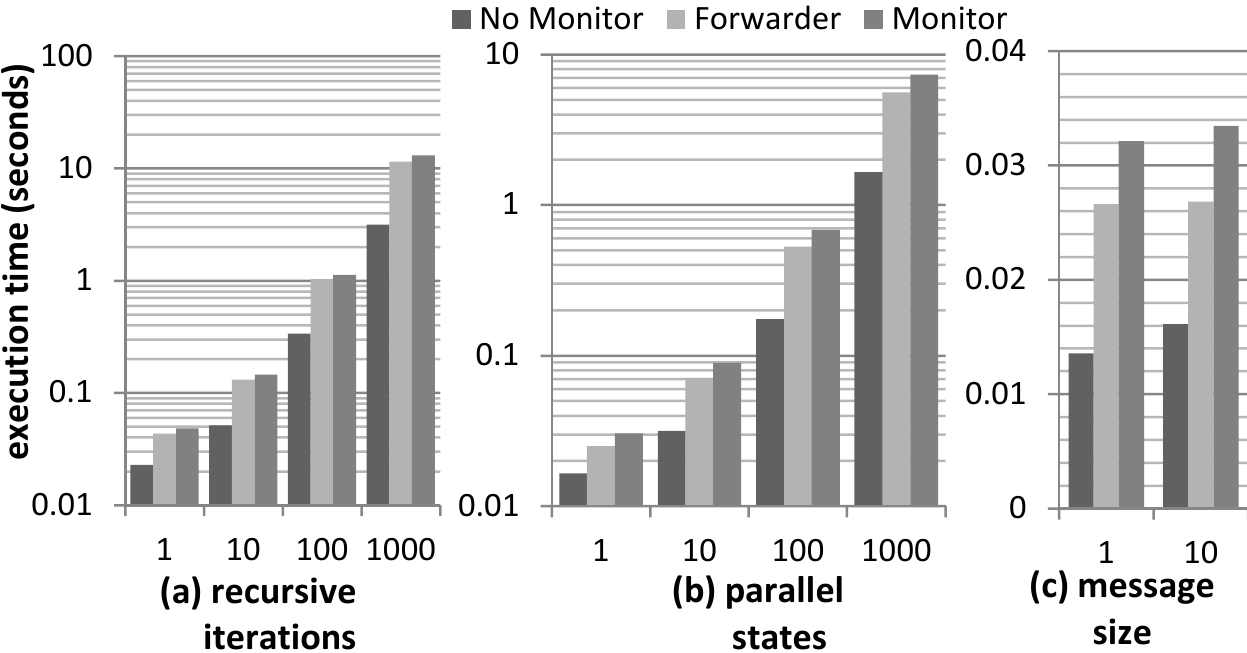}
\caption{Microbenchmarks comparing end-to-end monitor performance\label{fig:monitor:benchmarks}}
\end{figure}

\bparagraph{Benchmark environment and results}
The server and client endpoint processes, both monitors and the
RabbitMQ broker (2.7.0/R13B03) are all run on separate machines with
the same specification: Intel Core2 Duo 2.80~GHz and 4~GB main memory,
running 64-bit Ubuntu~11.04 (kernel~2.6.38) and connected via gigabit
Ethernet. Latency between each node is measured to be 0.24~ms on
average (ping 64 bytes). The benchmark applications are executed using
Python~2.7.1.

Figure \ref{fig:monitor:benchmarks} presents the results for the three
benchmark scenarios. Each chart gives the mean time (y-axis) for the
client and server to complete one session after repeating the
benchmark 100 times for each parameter configuration (session
length/parallel states/message size). Scenario~(3) message size is
measured for session length 1.
For all three scenarios, the results show that the overhead of the
monitor due to FSM generation and FSM-based message checking, the
baseline cost in the current framework, are acceptable (around 20\%).
Non-communication related computation in more
realistic applications and higher latency environments will both
contribute to decreasing the relative overhead. For scenario~(1) in
chart~(a), note that the relative overhead decreases (from 12\% to
9\%) as the session length increases, because the one-time FSM
generation cost becomes less prominent. Although our implementation
work is ongoing, we believe these results confirm the feasibility of
our approach.
As expected, the forwarding configuration incurs extra latencies (due
to the reciprocal shape of the benchmark protocol) in comparison to
the (No Monitor) case.
The full source code and raw results
of these benchmarks, and additional tests using protocols with assertions, can be obtained from the
project homepage \cite{spy}.
\section{Related Work}
\label{sec:related}
The work closest to ours is that by Ancona et al.~\cite{dalt12}. It explores session types protocols as a test framework for multiagent systems (MAS). A global session type is specified as cyclic Prolog terms in Jason (a MAS development platform) and verified through test monitors. Their global types are less expressive in comparison with the language presented in this paper (due to restricted arity on forks and the lack of assertions). Their monitor is centralised and 
global safety properties are not discussed. Kruger et al.~\cite{DBLP:journals/logcom/KrugerMM10} propose a run-time monitoring framework,
projecting MSCs to FSM-based distributed monitors. They use
aspect-oriented programming techniques to inject monitors into the
implementation of the components. Our outline monitoring verifies conversation protocols and does not require such monitoring-specific augmentation of programs. Gan \cite{Gan:2007} follows a similar but centralised approach to Kruger et al.

Works on monitoring BPEL languages can also be compared.
Baresi et al.~\cite{Baresi:2004} develop a run-time monitoring tool with assertions.
However, a major difference is that BPEL approaches do not treat
or prove global safety. BPEL is expressive, but does not support
distribution and is designed to work in a centralised manner.
\section{Conclusion and Future Work}
We have shown that session types are amendable for dynamic verification. Our implementation automates distributed monitoring by generating FSMs from local protocol projections. Further benchmarks are needed to compare the conversation API with existing network libraries and to investigate its performance. Future work includes also the incorporation of more elaborate handling of error cases into monitor functionality, extending Scribble and automatic generation of services stubs. 
Although our implementation work is ongoing, the results confirm the feasibility of
our approach. We believe this work is an important step towards a better, safer world of easier to speak and easier to understand distributed conversations.   

\textbf{Acknowledgments.} I would like to dedicate this paper to the memory of Kohei Honda, who is a constant source of inspiration to me and whose guidance was invaluable. I thank my supervisor Nobuko Yoshida for her constant support and ideas, my colleagues Raymond Hu and Pierre-Malo Deniélou for the discussions about the framework; the anonymous reviewers for useful comments and corrections; and Tzu-Chun Chen for her valuable feedback and her inspirational work on the formal system behind the presented work. This work is partially supported by VMWare PhD studentship and EPSRC EP/G015635/1.
\label{sec:conclusion}
{
\bibliographystyle{eptcs}
\bibliography{session}
}
\end{document}